\documentclass[pre,twocolumn,noshowpacs,showkeys,amsfonts,amssymb,amsmath,eqsecnum,letterpaper,nofootinbib,floatfix]{revtex4}

\usepackage{graphicx}
\begin{document}
\renewcommand{\r}{\mathbf{r}}
\newcommand{\deltaw}{\tilde{\delta}}
\newlength{\GraphicsWidth}
\setlength{\GraphicsWidth}{8cm}

\title{Confined Coulomb systems with adsorbing boundaries:\\
the two-dimensional two-component plasma} 
\author{Lina Merch\'an}
\email{l-mercha@uniandes.edu.co}
\author{Gabriel T\'ellez}
\email{gtellez@uniandes.edu.co} 
\affiliation{Departamento de F\'{\i}sica, Universidad de Los Andes,
A.A.~4976, Bogot\'a, Colombia}
\begin{abstract}
Using a solvable model, the two-dimensional two-component plasma, we
study a Coulomb gas confined in a disk and in an annulus with
boundaries that can adsorb some of the negative particles of the
system. We obtain explicit analytic expressions for the grand
potential, the pressure and the density profiles of the system. By
studying the behavior of the disjoining pressure we find that without
the adsorbing boundaries the system is naturally unstable, while with
attractive boundaries the system is stable because of a positive
contribution from the surface tension to the disjoining pressure. The
results for the density profiles show the formation of a positive
layer near the boundary that screens the adsorbed negative particles,
a typical behavior in charged systems. We also compute the adsorbed
charge on the boundary and show that it satisfies a certain number of
relations, in particular an electro-neutrality sum rule.
\end{abstract}

\maketitle

\section{Introduction}

In this paper we study the classical (i.e.~non-quantum) equilibrium
statistical mechanic properties of confined Coulomb systems with
adsorbing boundaries. A Coulomb system is a system of charged
particles interacting through the Coulomb potential. There are several
interesting realizations of Coulomb systems with several applications
such as plasmas, electrolytes, colloidal suspensions, etc... In the
present paper we are interested in the case where the Coulomb system
is confined with boundaries that can attract and adsorb some particles
of the system. Our study of this kind of systems will be done using a
solvable model of Coulomb system: the symmetric two-dimensional
two-component plasma, a system of two kind of oppositely charged
particles $\pm q$ at thermal equilibrium at an inverse temperature
$\beta=(k_B T)^{-1}$. The classical equilibrium statistical mechanics
of the system can be exactly solved when $\beta q^2=2$.

One can think of several examples where the present situation of
Coulomb systems confined with adsorbing boundaries is relevant, for
instance in a plasma or an electrolyte near an electrode with
adsorbing sites~\cite{RosinbergLebowitzBlum,Cornu-adsorption}. Another
situation in which we will focus our attention is in solutions of
amphiphathic molecules and ions for example in soap films and
bubbles. Amphiphathic molecules have an hydrophobic tail and an
hydrophilic charged head (usually negative) and, for this reason, when
they are submerged in water they rearrange themselves in such a way as
to minimize the contact of the hydrophobic tails with the surrounding
medium. They can achieve configurations such as bilayers, micelles and
vesicles, among others.

In a previous paper~\cite{TellezMerchan-jabon} we studied a soap film
by modeling it as a Coulomb system confined in a slab. A soap film can
be seen as a system of amphiphathic molecules in a bilayer
configuration with a water inner layer. The overall neutral system
with negatively charged amphiphathic anions and positive micro-cations
(usually Na${}^+$) in water was modeled as a two-dimensional
two-component plasma. The two dimensions were in the breadth of the
film not on the surface: we studied a cross section of the
film. Because of the hydrophobicity of their tails, the soap anions
prefer to be in the boundaries of the film. This was modeled by a
one-body attractive short-range external potential acting over
them. This means that the negative particles felt an attractive
potential over a small distance near each boundary. In this sense the
negative particles of the system can be ``adsorbed'' by the boundary.

In Ref.~\cite{TellezMerchan-jabon} we found exact expressions for the
density, correlations and pressure inside the film.  By studying the
disjoining pressure, we were able to conclude that the Coulomb
interaction plays an important role in the collapse of a thick soap
film to a much thinner film. Actually if a large number of amphiphathic
molecules are in the boundary due to a large strength of the
attractive external potential near the boundary, the system is always
stable. On the other hand if the attractive potential near the
boundary is not strong enough a thick film will not be stable and will
collapse to a thin film.

These results can be compared (qualitatively due to the simplicity of
the model under consideration) to the experimental situation of the
transition of a thick film to a thin black film. These black film
phenomena occur when the soap films width is smaller than visible
light wavelength and it is seen black. Two types of black films are
observed experimentally: the common black film and the much thinner
Newton black film.

It is interesting to know to what extent the results of our previous
work~\cite{TellezMerchan-jabon} depend on the geometry. Here we study
this two-component plasma system confined in two other geometries, a
disk and an annulus. As we mentioned before, amphiphathic molecules in
water can achieve micelles and vesicles among others
configurations. In two dimensions, a cylindrical micelle can be seen
as a disk and a cylindrical vesicle as an annulus. If the length of
the cylindrical micelle or vesicle if much larger that its radius it
is reasonable to assume that the system in invariant in the
longitudinal direction and so we study only a cross section of
the system: a disk or an annulus. The Coulomb interaction is then the
two-dimensional Coulomb potential which is $v_c(r)=-\ln (r/d)$ for two
particles at a distance $r$ of each other. The length $d$ is an
arbitrary length which fixes the zero of the potential.

Two-dimensional Coulomb systems with log interaction have properties
that are similar to those in three dimensional charged systems with
the usual $1/r$ potential. They satisfy Gauss law and Poisson equation
in two dimensions. Several universal properties, such as screening
effects, are direct consequences of the harmonic nature of the $-\ln
(r/d)$ and $1/r$ potentials, which are the solutions of the two- and
three-dimensional Poisson equation.  Therefore the exact solutions
obtained for the 2D models play an important role in understanding
real 3D Coulomb systems.

The rest of this work is organized as follows. In
section~\ref{sec:Model} we present in detail the system under
consideration and briefly review how this model can be exactly
solved. In section~\ref{sec:Pressure} we compute the grand potential
of the system and the disjoining pressure and study the stability of
the system. In section~\ref{sec:Density} we compute the density
profiles of the different types of particles in the system and the
adsorbed charge on the boundaries. Finally, we conclude recalling the
main results of the present work.

%%%%%%%%%%%%%%%%%%%%%%%%%%%%%%%%%%%%%%%%%%%%%%%%%%%%%%%%%%%%%%%%%%%

\section{The Model and Method of Solution}
\label{sec:Model}

The system under consideration is a two-dimensional system composed of
two types of point particles with charges $\pm q$. Two particles with
charges $sq$ and $s'q$ at a distance $r$ apart interact with the
two-dimensional Coulomb potential $-ss'q^2 \ln(r/d)$ where $d$ is an
arbitrary length. This system is known as the symmetric
two-dimensional two-component plasma.

When the Boltzmann factor for the Coulomb potential is written, the
adimensional coulombic coupling constant appears: $\Gamma=q^2/k_B
T=\beta q^2$. Notice that in two dimensions $q^2$ has dimensions of
energy. For a system of point particles if $\Gamma\geq 2$ the system
is unstable against the collapse of particles of opposite sign, the
thermodynamics of the system are not well defined unless one considers
hard-core particles or another regularization procedure (for instance
a lattice model instead of a continuous gas). On the other hand, if
$\Gamma<2$ the thermal agitation is enough to avoid the collapse and
the system of point particles is well defined. The two-component
plasma is known to be equivalent to the sine-Gordon model and using
this relationship and the results known for this integrable field
theory, the thermodynamic properties of the two-component plasma in
the bulk have been exactly determined~\cite{SamajTravenec} in the
whole range of stability $\Gamma<2$. However, there are no exact
results for confined systems for arbitrary $\Gamma$ (with the
exception of a Coulomb system near an infinite plane
conductor~\cite{JancoSamaj-metal} or ideal
dielectric~\cite{Samaj-ideal-diel} wall).

It is also well-known for some time that when $\Gamma=2$ the
sine-Gordon field theory is at its free fermion point. This means that
the system is equivalent to a free fermion field theory and therefore
much more information on the system can be obtained. In particular the
thermodynamic properties and correlation functions can be exactly
computed even for confined systems in several different geometries and
different boundary
conditions~\cite{Gaudin,CornuJanco-CGas,CornuJanco,Forrester-metal-tcp,%
JanManPis,JancoTellez-coulcrit,JancoSamaj-diel,Tellez-tcp-neumann}. From
now on we will consider only the case when $\Gamma=2$.

Since at $\Gamma=2$ a system of point particles is not stable one
should start with a regularized model with a cutoff distance $a$ which
can be the diameter of the hard-core particles or the lattice spacing
in a lattice model~\cite{Gaudin}. The system is worked out in the
grand-canonical ensemble at given chemical potentials $\mu_{+}$ and
$\mu_{-}$ for the positive and negative particles respectively. In the
limit of a continuous model $a\to0$ the grand partition function and
the bulk densities diverge. However the correlation functions have a
well-defined limit. In this continuous limit it is useful to work with
the rescaled fugacities~\cite{CornuJanco} $m_{\pm}=2\pi d
e^{\beta\mu_{\pm}}/a^2$ that have inverse length dimensions. The
length $m^{-1}$ can be shown to be the screening length of the
system~\cite{CornuJanco}.  If external potentials $V_{\pm}(\r)$ act on
the particles (as in our case) it is useful to define position
dependent fugacities $m_{\pm}(\r)=m_{\pm} \exp(-\beta V_{\pm}(\r))$.

Let us briefly review the method of resolution described by Cornu and
Jancovici \cite{CornuJanco-CGas,CornuJanco} for the
two-component plasma.  It will be useful to use the complex
coordinates $z=re^{i\theta}=x + i y$ for the position of the
particles. For a continuous model, $a\to0$, ignoring the possible
divergences for the time being, it is shown in Ref.~\cite{CornuJanco}
that the equivalence of the two-component plasma with a free fermion
theory allows the grand partition function to be written as
\begin{equation}
\label{Xi_matrix}
\Xi =\det \left[ \left( \begin{array}{cc}
0 & 2\partial_z\\
2\partial_{\bar{z}} & 0
\end{array}\right)^{-1} \left( \begin{array}{cc}
m_{+}(\r) & 2\partial_z\\
2\partial_{\bar{z}} & m_{-}(\r)
\end{array}\right) \right]
\end{equation}
Then defining an operator $K$ as
\begin{equation}
\label{eq:def-K}
K = \left( \begin{array}{cc}
0 & 2\partial_z\\
2\partial_{\bar{z}} & 0
\end{array}\right)^{-1} \left( \begin{array}{cc}
m_{+}(\r) & 0\\
0 & m_{-}(\r)
\end{array}\right)
\end{equation}
the grand partition function $\Xi$ can be expressed as
\begin{equation}
\label{grand_partition}
\Xi=\det \left( 1 + K\right).
\end{equation}
The calculation of the grand potential $\beta\Omega=-\ln \Xi$ and the
pressure $p=-\partial \Omega/\partial V$ where V is the volume (in a
two-dimensional system this refers to the area), reduces to finding
the eigenvalues of $K$ because the grand potential can be written as
\begin{equation}
\Omega=-k_B T\sum_{i}\ln (1+\lambda_i)
\end{equation} 
where $\lambda_i$ are the eigenvalues of the operator K. 

On the other hand, the calculation of the one-particle densities and
correlations reduces to finding a special set of Green functions. As
usual, the density can be expressed as a functional derivative
\begin{equation}
\rho_{\pm}(r)=m_{\pm}(r)\frac{\delta \ln \Xi}{\delta m_{\pm}(r)}
\end{equation}
If we define the $2\times 2$ matrix 
\begin{equation}
\mathbf{G}(\r_1,\r_2)=\left(
\begin{array}{cc}
  G_{++}(\r_1,\r_2) &   G_{+-}(\r_1,\r_2) \\
  G_{-+}(\r_1,\r_2) &   G_{--}(\r_1,\r_2) 
\end{array}
\right)
\end{equation}
 as the kernel of the inverse of the operator
\begin{equation}
\left(\begin{array}{cc}
m_{+}(\r) & 2\partial_z\\
2\partial_{\bar{z}} & m_{-}(\r)
\end{array}
\right)
\end{equation}
then the one-body density and two-body Ursell functions can be
expressed in terms of these Green functions as
\begin{eqnarray}
\label{rho_Green}
\rho_{s_1}(\r_{1})&=& m_{s_1} G_{s_1 s_1}\left(\r_{1}, \r_{1}\right),
\nonumber\\ 
\rho_{s_1 s_2}^{(2)T}\left( \r_{1},
\r_{2}\right)&=&-m_{s_1}m_{s_2}G_{s_1 s_2}\left( \r_{1},
\r_{2}\right)G_{s_2 s_1}\left( \r_{2}, \r_{1}\right)\nonumber\\
\end{eqnarray}
where $s_{1,2}\in\{+,-\}$ denote the sign of the particles.  In polar
coordinates these Green functions satisfy the following set of
equations
\begin{equation}
\label{eq:GreenFunctions-G}
\left[
\begin{array}{cc}
m_+(\r_1)&e^{-i\theta_1}\left[\partial_{r_1}-\frac{i
\partial_{\theta_1}}{r_1}\right]\\
e^{i\theta_1}\left[\partial_{r_1}+\frac{i
\partial_{\theta_1}}{r_1}\right]&m_-(\r_1)
\end{array}
\right]
\mathbf{G}=\delta(\r_1-\r_2)\mathbb{I}
\end{equation}
with $\mathbb{I}$ being the unit $2\times2$ matrix.

The above formalism is very general, it can be applied to a variety of
situations. In the case we are interested in, we will consider two
geometries in which the system is confined: a disk of radius $R$ and
an annulus of inner and outer radius $R_1$ and $R_2$ respectively.

The negative particles are supposed to model amphiphathic molecules and
therefore they are attracted to the boundaries of the system while
positive particles are not. This is modeled by an attractive external
one-body potential $V_{-}(\r)$ acting on the negative particles near
the boundary while for the positive particles $V_{+}(\r)=0$.

Actually we will consider two models for this potential. In the first
model (model I) the fugacity $m_{-}(\r)$ for the negative particles
reads, for the disk geometry,
\begin{equation}\label{eq:m-disk}
m_{-}(\r)=m e^{-\beta V_{-}(\r)}=m+\alpha\delta(r-R)
\end{equation}
inside the disk, while $m_{+}=m$ is constant. Outside the disk $r>R_2$
both fugacities vanish. In the annulus geometry,
\begin{equation}\label{eq:m-annulus}
m_{-}(\r)=m+\alpha_1\delta(r-R_1)+\alpha_2
\delta(r-R_2)
\end{equation}
inside the annulus. The coefficients $\alpha$, $\alpha_{1}$ and
$\alpha_{2}$ measure the strength of the attraction to the walls. In
the following we will call these coefficients adhesivities.

In the second model (model II) that we will eventually consider the
external potential $V_{-}(\r)$ is a step function with a range
$\deltaw$ of attraction near the boundary. This model allows us to
obtain valuable information regarding the frontier regions. Actually
we will report here only the main results for the annulus geometry
with model II in section~\ref{sec:modelII}, further results on
this model on the disk geometry can be found in
Ref.~\cite{Lina-tesis}.

In the two following sections we will apply the method presented here
to obtain the grand potential and the density profiles of the system.

%%%%%%%%%%%%%%%%%%%%%%%%%%%%%%%%%%%%%%%%%%%%%%

\section{The pressure}
\label{sec:Pressure}

\subsection{The grand potential}
To find the pressure of the confined Coulomb system we proceed first
to compute the grand potential. As shown in section~{\ref{sec:Model}}
the grand potential $\Omega$ is given by
\begin{equation}
\label{eq:Omega-prod-lambda}
  \beta \Omega = - \ln \prod_i(1+\lambda_i)
\end{equation}
where $\lambda_i$ are the eigenvalues of the operator $K$ given by
equation~(\ref{eq:def-K}). The eigenvalue problem for $K$ with
eigenvalue $\lambda$ and eigenvector $(\psi,\chi)$ reads
\begin{subequations}
\label{eq:vp-K}
\begin{eqnarray}
\label{eq:vp-K-1}
  m_{-}(\r) \chi(\r) &=& 2\lambda\partial_{\bar{z}} \psi(\r)\\
\label{eq:vp-K-2}
  m_{+}(\r) \psi(\r) &=& 2\lambda\partial_{z} \chi(\r)
\end{eqnarray}
\end{subequations}
These two equations can be combined into
\begin{equation}
  \Delta \chi(\r) = \frac{m_{+}(\r) m_{-}(\r)}{\lambda^2}\ \chi(\r)
\end{equation}
We now detail the computation of the grand potential for the case of a
Coulomb system confined in a disk. The annulus geometry follows
similar calculations.

We will only use model I for the attracting potential on the boundary
for the calculation of the grand potential. Using model I the
fugacities inside the disk read
\begin{equation}
  m_+(\r)=m\,, \qquad m_{-}(\r)= m + \alpha \delta(r-R)\,,
\end{equation}
and outside the disk they vanish.  Replacing these fugacities into the
eigenvalue problem equations~(\ref{eq:vp-K}) we find that $\chi(\r)$
is a continuous function while $\psi(\r)$ is discontinuous at $r=R$
due to the Dirac delta distribution in $m_{-}(\r)$. The discontinuity
of $\psi$ is given by
\begin{equation}
\label{eq:discont-psi}
  \psi(R^{+},\theta)-\psi(R^{-},\theta)=\frac{\alpha}{\lambda}
  \chi(R,\theta)\,e^{-i\theta}
\end{equation}
Defining $k=m/\lambda$, inside the disk $r<R$, $\chi(\r)$ obeys
the equation
\begin{equation}
  \Delta \chi(\r) = k^2 \chi(\r)
\end{equation}
with solutions of the form
\begin{equation}
  \chi(r,\theta)=A_l e^{il\theta} I_{l}(kr)
\end{equation}
and
\begin{equation}
  \psi(r,\theta)=A_l e^{i(l-1)\theta} I_{l-1}(kr) 
\end{equation}
where $I_l$ is a modified Bessel function of order $l$. Outside the
disk $m_-=m_+=0$ and the corresponding solutions to
equations~(\ref{eq:vp-K}) are that $\psi$ is analytic and $\chi$
anti-analytic, namely,
\begin{eqnarray}
  \chi(\r)&=&B_l e^{il\theta} r^{-l}\, \\
  \psi(\r)&=&B_{l} e^{i(l-1)\theta} r^{l-1}\,.
\end{eqnarray}
In order to have vanishing solutions at $r\to\infty$ it is necessary
that
\begin{subequations}
\begin{eqnarray}
  \psi(R^{+},\theta)=0, \quad \text{if\ } l\geq 1\\
  \chi(R,\theta)=0, \quad \text{if\ } l\leq 0
\end{eqnarray}
\end{subequations}
Using these boundary conditions together with the continuity of $\chi$
at $R$ and the discontinuity~(\ref{eq:discont-psi}) of $\psi$ at $R$
gives the eigenvalue equation
\begin{subequations}
\label{eq:eigenval-eq}
\begin{eqnarray}
 I_l\left(\frac{mR}{\lambda}\right)&=&0\quad\text{if\ }l\leq 0\\
\alpha I_l\left(\frac{mR}{\lambda}\right)
+I_{l-1}\left(\frac{mR}{\lambda}\right)
&=&0\quad\text{if\ }l\geq 1 
\end{eqnarray}
\end{subequations}
The product appearing in equation~(\ref{eq:Omega-prod-lambda}) can be
partially computed by recognizing that the l.h.s of the eigenvalue
equation~(\ref{eq:eigenval-eq}) for arbitrary $\lambda$ can be written
as a Weierstrass
product~\cite{Forrester,JancoTellez-coulcrit,Tellez-tcp-neumann,TellezMerchan-jabon}.
Let us introduce the analytic functions
\begin{eqnarray}
  f_l^{(-)}(z)&=&I_l(mzR) \ l! \left(\frac{2}{mzR}\right)^l \\
  f_l^{(+)}(z)&=& \left[
\alpha I_l(mzR)+I_{l-1}(mzR)\right] 
(l-1)!
\left(\frac{2}{mzR}\right)^{l-1}\nonumber \\
\end{eqnarray}
By construction the zeros of $f^{(+)}_{l}$ are the inverse of the
eigenvalues $\lambda$ for $l>0$ and the zeros of $f^{(-)}_{-l}$ are
the inverse eigenvalues $\lambda$ for $l\leq0$. Furthermore, since
$f_l^{(\pm)}(0)=1$, ${f'}_l^{(\pm)}(0)=0$ and $f_l^{(\pm)}(z)$ is an even
function it can be factorized as the Weierstrass product
\begin{equation}
  f_l^{(\pm)}(z)=\prod_{\lambda_l}
  \left(1-\frac{z}{\lambda_{l}^{-1}}\right)
\end{equation}
where the product runs over all $\lambda_l$ solutions of
equation~(\ref{eq:eigenval-eq}) for a given $l$.  Then we can conclude
that the grand potential~(\ref{eq:Omega-prod-lambda}) is given by
\begin{equation}
  \beta\Omega=\sum_{l=0}^{\infty} \ln f_l^{(-)}(-1)
+\sum_{l=1}^{\infty} \ln f_l^{(+)}(-1)
\end{equation}
After shifting by one the index in the second sum and rearranging the
expression we find the final result for the grand potential
\begin{equation}
  \label{eq:grand-pot-disk-result}
\Omega^{\text{D}}=\Omega_{\text{hw}}^{\text{D}}+
\Omega_{\text{at}}^{\text{D}}
\end{equation}
with
\begin{equation}
  \label{eq:grand-pot-hard-disk}
\beta\Omega_{\text{hw}}^{\text{D}}=-2\sum_{l=0}^{\infty}
\ln\left[
l!\left(\frac{2}{mR}\right)^l I_l(mR)\right]
\end{equation}
which is the grand potential for a two-component plasma in a disk with
hard wall boundaries~\cite{JanManPis} ($\alpha=0$) and
\begin{equation}
  \label{eq:grand-pot-disk-attract}
\beta\Omega_{\text{at}}^{\text{D}}
=-\sum_{l=0}^{\infty}
\ln\left[
1+\alpha\, \frac{I_{l+1}(mR)}{I_l(mR)}\right]
\end{equation}
is the contribution due to the attractive potential near the walls.

Now we turn our attention to the case of the Coulomb system confined
in an annulus of inner radius $R_1$ and outer radius $R_2$. In this
case the position-dependent fugacity for the negative particles inside
the annulus is given by
\begin{equation}
  m_{-}(\r)=m+\alpha_1 \delta(r-R_1) +\alpha_2 \delta(r-R_2)
\end{equation}
The calculation of the grand potential follows similar steps as
above. One should solve the Laplacian eigenvalue problem with the
appropriate boundary conditions given by the continuity of $\chi$ and
the discontinuity of $\psi$ at $R_1$ and $R_2$. After some
straightforward calculations the final result for the grand potential
is
\begin{equation}
\Omega^{\text{A}}=\Omega_{\text{hw}}^{\text{A}}+
\Omega_{\text{at}}^{\text{A}}
\end{equation}
with
\begin{widetext}
\begin{equation}
\label{eq:grand-pot-har-annulus}
  \beta\Omega_{\text{hw}}^{\text{A}}=-2\sum_{l=0}^{\infty}
\ln\left[
\frac{mR_1^{l+1}}{R_2^l}
\left(
I_l(mR_2)K_{l+1}(mR_1)+
I_{l+1}(mR_1)K_l(mR_2)
\right)
\right]
\end{equation}
and
\begin{eqnarray}
  \label{eq:grand-pot-annulus-attract}
  \beta\Omega_{\text{at}}^{\text{A}}&=&
  -\sum_{l=0}^{\infty} \ln\left[1+\alpha_1
    \frac{I_l(mR_2)K_l(mR_1)-K_l(mR_2)I_l(mR_1)}{I_l(mR_2)K_{l+1}(mR_1)+
      I_{l+1}(mR_1)K_l(mR_2)}\right]\nonumber\\
  & & -\sum_{l=0}^{\infty} \ln\left[1+\alpha_2
    \frac{I_{l+1}(mR_2)K_{l+1}(mR_1)-K_{l+1}(mR_2)I_{l+1}(mR_1)}{I_l(mR_2)
      K_{l+1}(mR_1)+I_{l+1}(mR_1)K_l(mR_2)}
    \right]
\end{eqnarray}  
\end{widetext}
The first term $\Omega_{\text{hw}}^{\text{A}}$ is the grand potential
for a two-dimensional two-component plasma confined in an
annulus~\cite{JanManPis} with hard wall boundaries\footnote{
Eq.~(4.16) of Ref.~\cite{JanManPis} for the grand potential in an
annulus with hard wall boundaries is incorrect, however the equation
above~(4.16) is correct and gives our
result~(\ref{eq:grand-pot-har-annulus}) for the grand potential }
($\alpha=0$) and the second term $\Omega_{\text{at}}^{\text{A}}$ is
the contribution to the grand potential due to the attractive nature
of the walls.

It should be noted that all sums~(\ref{eq:grand-pot-hard-disk}),
(\ref{eq:grand-pot-disk-attract}), (\ref{eq:grand-pot-har-annulus})
and~(\ref{eq:grand-pot-annulus-attract}) above are divergent and
should be cutoff to obtain finite results. This is due to the fact
that the two-component plasma of point particles is not stable against
the collapse of particles of opposite sign for $\beta q^2 \geq 2$ and
a short-distance cutoff $a$ should be introduced, $a$ can be
interpreted as the hard-core diameter of the particles.  If $R$ is the
characteristic size of the system (for instance the radius of the disk
in the disk case) then $l/R$ is a wave-length and the short-distance
cutoff $a$ gives an ultraviolet cutoff $1/a$. Then the cutoff for $l$
(say $N$) in the sums should be chosen of order
$R/a$~\cite{CornuJanco,JanManPis}.

\subsection{Finite-size corrections}

It is instructive to study the behavior of the grand potential when
the system is large. It has been known for some time that
two-dimensional Coulomb systems in their conducting phase have a
similar behavior to critical
systems~\cite{JanManPis,JancoTellez-coulcrit}. In particular the grand
potential of a two-dimensional Coulomb system confined in a domain of
characteristic size $L$ has a large-$L$ expansion
\begin{equation}\label{eq:Finite-Size-Corrections}
\beta\Omega= A L^2 + B L + \frac{\chi}{6} \ln L + O(1)
\end{equation}
similar to the one predicted by Cardy~\cite{CardyPeschel,Cardy} for
critical systems. The first two terms are respectively the bulk grand
potential and the surface contribution to the grand potential (the
surface tension) and are non-universal. The logarithmic term is a
universal finite-size correction to the grand potential, it does not
depend on the microscopic detail of the system, only on the topology
of the manifold where the system lives through the Euler
characteristic $\chi$. For a disk $\chi=1$ and for an annulus
$\chi=0$.

It is interesting to verify if this finite-size expansion holds for
the systems studied here, in particular if the finite-size correction
is modified by the special attractive nature of the walls considered
here.

Let us first consider the case of the disk geometry. We choose to
cutoff the sums~(\ref{eq:grand-pot-hard-disk})
and~(\ref{eq:grand-pot-disk-attract}) to a maximum value for $l$ equal
to $R/a$ and the results given here are for $a\to0$. The
finite-size expansion for large-$R$ of the hard wall contribution to
the grand potential has already been computed in Ref.~\cite{JanManPis}
with the result
\begin{equation}
  \beta\Omega_{\text{hw}}^{\text{D}}= -\beta p_{b} \pi R^2 +
\beta\gamma_{\text{hw}} 2\pi R + \frac{1}{6}\ln(mR)+O(1)
\end{equation}
where the bulk pressure $p_b$ is given by
\begin{equation}
  \label{eq:bulk-pressure}
  \beta p_b=\frac{m^2}{2\pi}\left(1+\ln\frac{2}{ma}\right)
\end{equation}
and the surface tension $\gamma_{\text{hw}}$ for hard walls is given by
\begin{equation}
  \beta \gamma_{\text{hw}}=m\left(\frac{1}{4}-\frac{1}{2\pi}\right) 
\end{equation}
We only need to compute the large-$R$ expansion
of~(\ref{eq:grand-pot-disk-attract}). This can be done expressing the
Bessel function $I_{l+1}(mR)$ as $I_{l+1}(mR)= I'_l(mR)+ l I_l(mR)
/(mR)$, using the uniform Debye expansions~\cite{AbramowitzStegun} of
the Bessel functions valid for large argument
\begin{subequations}
\label{eq:Debye-exp-BesselI}
\begin{eqnarray}
  I_{l}(z) &\hskip-7pt\sim&\hskip-7pt 
  \frac{e^\eta}{\sqrt{2\pi}(l^2+z^2)^{1/4}} \left[1+\frac{3 t- 5 t^3}{24
  l}\hskip-1pt+\hskip-1pt\cdots\right]\\ 
  I_{l}'(z)%&\hskip-6pt\sim&\hskip-6pt 
  &\sim&\frac{(l^2+z^2)^{1/4}
  e^{\eta}}{\sqrt{2\pi}\,z} \left[1-\frac{9 t- 7t^3}{24 l}+\cdots\right]
\end{eqnarray}
\end{subequations}
with $\eta=\sqrt{l^2+z^2}+l\ln\left(\frac{z}{l+\sqrt{l^2+z^2}}\right)$ and
$t=l/\sqrt{l^2+z^2}$, and using the Euler-McLaurin formula to
transform the summation into an integral
\begin{eqnarray}
  \sum_{l=0}^N f(l)&=&\int_0^N
  f(x)\,dx+\frac{1}{2}\left[f(N)+f(0)\right]\\
&&+\frac{1}{12}\left[f'(N)-f'(0)\right]+\cdots\nonumber
\end{eqnarray}
After some calculations, taking first the limit $N\to\infty$, keeping
only the non-vanishing terms, then taking the limit $R\to\infty$ and
replacing $N$ by $R/a$, we find that~(\ref{eq:grand-pot-disk-attract})
contributes only to the surface tension giving for the grand potential
the final result
\begin{equation}\label{eq:gd-pot-finite-size-exp-disk}
  \beta\Omega^{\text{D}}= -\beta p_{b} \pi R^2 + 2\pi R \beta\gamma+
  \frac{1}{6}\ln(mR)+O(1)
\end{equation}
where the surface tension is now given by
\begin{equation}
  \label{eq:surface-tension}
  \beta\gamma=-\frac{m}{4\pi}
\left[
\alpha\ln\frac{2}{ma}+1-\pi+\alpha
+\frac{1-\alpha^2}{\alpha}
\ln(\alpha+1)\right]
\end{equation}
We recover as expected the surface tension obtained in
Ref.~\cite{TellezMerchan-jabon} for the same system but confined in a
slab.

The universal logarithmic finite-size correction $(1/6)\ln (mR)$ is
still present and it is not modified by the presence of the attractive
boundaries.

For the annulus geometry we are interested in the limit $R_1 \to
\infty$ and $R_2\to\infty$ with $R_2/R_1$ finite. We proceed as above,
using also this time the Debye expansion for the Bessel
functions~\cite{AbramowitzStegun}
\begin{subequations}
\label{eq:Debye-exp-BesselK}
\begin{eqnarray}
  K_{l}(z) &\hskip-7pt\sim&\hskip-7pt
  \frac{\sqrt{\pi}e^{-\eta}}{\sqrt{2}(l^2+z^2)^{1/4}} \left[1-\frac{3
  t- 5 t^3}{24 l}\hskip-1pt+\hskip-1pt\cdots\right]\\
  K_{l}'(z)&\hskip-6pt\sim&\hskip-6pt \frac{-\sqrt{\pi}(l^2+z^2)^{1/4}
  e^{-\eta}}{\sqrt{2}\,z} \left[1+\frac{9 t- 7 t^3}{24
  l}+\cdots\right]\nonumber\\
\end{eqnarray}
\end{subequations}
Using~(\ref{eq:Debye-exp-BesselI}) and~(\ref{eq:Debye-exp-BesselK})
one can notice that inside the logarithm
in~(\ref{eq:grand-pot-har-annulus}) the term $I_{l+1}(mR_1)
K_{l}(mR_2)$ is exponentially small compared to
$I_{l}(mR_2)K_{l+1}(mR_1)$ since $R_2-R_1\to\infty$. Also in the
contribution from the attractive
boundaries~(\ref{eq:grand-pot-annulus-attract}) the dominant terms are
\begin{eqnarray}
\label{eq:dominant-surface}
  \beta\Omega_{\text{at}}^{\text{A}}&\sim&
-\sum_{l=0}^{N}
\ln\left[1+\alpha_1\frac{K_{l}(mR_1)}{K_{l+1}(mR_1)}\right]
\nonumber\\
&& -
\sum_{l=0}^{N}
\ln\left[1+\alpha_2\frac{I_{l+1}(mR_2)}{I_{l}(mR_2)}\right]
\end{eqnarray}
Then, after some calculations we get the final result
\begin{equation}\label{eq:annulus-finite-size-exp}
  \beta\Omega^{\text{A}}=-\beta p_b \pi(R_2^2-R_1^2)
+\beta\gamma_1 2\pi R_1+\beta\gamma_2 2\pi R_2+O(1)
\end{equation}
The surface tension for each boundary is given by
\begin{equation}
  \beta\gamma_{i}= -\frac{m}{4\pi}
\left[
\alpha_{i}\ln\frac{2N}{R_{i}}+1-\pi+\alpha_{i}
+\frac{1-\alpha_{i}^2}{\alpha_{i}}
\ln(\alpha_{i}+1)\right]
\end{equation}
where $i=1$ for the inner boundary and $i=2$ for the outer boundary.
The cutoff $N$ should be chosen~\cite{JanManPis} as $R/a$ where here
$R=R_2 x^{-x^2/(1-x^2)}$ with $x=R_1/R_2$ in order to insure
extensivity and recover for the bulk pressure $p_b$ the same
expression~(\ref{eq:bulk-pressure}) as before. 

In the limit $a\to 0$ the leading term of the surface tension is the
same as in the slab and disk geometry
\begin{equation}\label{eq:surface-tension-domin}
  \beta\gamma\sim -\frac{\alpha m}{4\pi} \ln \frac{2}{ma}
\end{equation}
for a wall with adhesivity $\alpha$.

In equation~(\ref{eq:annulus-finite-size-exp}) we do not find any
logarithmic finite-size correction. There are some terms of the form
$\ln R_2/R_1$ which are order 1 because $R_2/R_1$ is finite. This is
in accordance with the expected
formula~(\ref{eq:Finite-Size-Corrections}) for an annulus where the
Euler characteristic is $\chi=0$. Here again the special attractive nature of
the walls does not modify the universal finite-size correction. 

As a conclusion to this part we might say that the logarithmic
correction is really universal, not only it does not depend on the
microscopic constitution of the system but also it is insensitive to
the existence of a short-range one-body potential near the walls.

\subsection{The disjoining pressure}

Let us first detail the case of the disk geometry. The pressure $p$ is
given in terms of the grand potential $\Omega$ by
\begin{equation}
p=-\frac{1}{2\pi R}\frac{\partial\Omega}{\partial R}
\end{equation}
Using equation~(\ref{eq:grand-pot-disk-result}) together with
equations~(\ref{eq:grand-pot-hard-disk})
and~(\ref{eq:grand-pot-disk-attract}) gives
\begin{equation}\label{eq:pressure-disk-total}
  p=p_{\text{hd}}+p_{\text{at}}
\end{equation}
where $p_{\text{hd}}$ is the pressure for a disk with hard walls
boundaries given by
\begin{equation}\label{eq:pressure-hd}
  \beta p_{\text{hd}}=
  \frac{m}{\pi R} \sum_{l=0}^{\infty} \frac{I_{l+1}(mR)}{I_{l}(mR)}
\end{equation}
and $p_{\text{at}}$ is the contribution to the pressure due to the
attractive potential near the walls and it is given by
\begin{widetext}
\begin{equation}\label{eq:pressure-atd}
  \beta p_{\text{at}}=
  \frac{\alpha}{2\pi R^2}
  \sum_{l=0}^{\infty}
  \frac{mR\left[I_l^2(mR)-I_{l+1}^2(mR)\right]-(2l+1)I_{l+1}(mR)
  I_l(mR)}{I_l(mR) \left[I_l(mR)+\alpha I_{l+1}(mR)\right]}
\end{equation}
\end{widetext}

To study the stability of the system against an external applied
pressure one should study the disjoining pressure $p_d=p-p_b$ defined
as the difference between the pressure of the system and the pressure of an
infinite system (the bulk pressure). Let us first consider the case
$\alpha=0$. A proper way to subtract the bulk pressure from
expression~(\ref{eq:pressure-hd}) is by using the equation of state of
the infinite system~\cite{CornuJanco}
\begin{equation}
  \beta p_b= \frac{1}{2} n_b + \frac{m^2}{4\pi}
\end{equation}
where $n_b$ is the bulk total density. A simple scaling argument shows
that for $\beta q^2<2$ the equation of state of the two-dimensional
two-component plasma is~\cite{HaugeHemmer}
\begin{equation}\label{eq:eq-of-state}
\beta p_b=[1-(\beta q^2/4)]n_b
\end{equation}
For $\beta q^2=2$, the case considered here, the introduction of a
cutoff $a$ is needed in order to avoid divergences but this
breaks the scale invariance of the two-dimensional logarithmic Coulomb
potential giving rise to the anomalous term $(m^2/4\pi)$ in the equation
of state. Notice that when the cutoff $a\to 0$ both $p_b$ and
$n_b$ diverge and we have $\beta p_d/n_b=1/2$ in accordance to the
general equation of state~(\ref{eq:eq-of-state}).

Formally, the bulk density can be written as (see next section for
details)
\begin{eqnarray}
  n_b&=&\frac{m^2}{\pi}\sum_{l=-\infty}^{+\infty} I_l(mR) K_l(mR)\nonumber\\
  &=&\frac{m^2}{\pi}\left[I_0 K_0+2\sum_{l=1}^{\infty} I_l K_l
  \right]
\end{eqnarray}
In the above expression and below the omitted argument of the Bessel
functions is $mR$ unless stated otherwise. On the other hand using the
Wronskian~\cite{Grad} $I_l K_{l+1} + I_{l+1} K_l = (1/mR)$ the hard disk
pressure~(\ref{eq:pressure-hd}) can be formally written as
\begin{equation}
  \beta p_{\text{hd}}= \frac{m^2}{\pi}\left[
    \sum_{l=0}^{\infty} \frac{I_{l+1}^2 K_l}{I_l} +
    \sum_{l=1}^{\infty} I_l K_l
    \right]
\end{equation}
Then the disjoining pressure for $\alpha=0$ is given by
\begin{equation}\label{eq:p-hd-disj}
\beta (p_{\text{hd}}-p_b)=
  \beta p_{\text{hd,disj}}=
  \frac{m^2}{\pi}\left[\sum_{l=0}^{\infty} \frac{I_{l+1}^2 K_l}{I_l} 
    -\frac{I_0 K_0}{2} - \frac{1}{4}
    \right]
\end{equation}
Although the pressure $p_{\text{hd}}$ and the bulk pressure $p_{b}$
are divergent when the cutoff $a$ vanishes, the disjoining pressure
$p_{\text{hd,disj}}$ in the case $\alpha=0$ proves to be well-defined
for $a\to 0$ and the series~(\ref{eq:p-hd-disj}) is convergent.

%%%%%%%%%%%%%%%%%%%%%%%%%%%%%%%%%%
%
%   Figure 1:
%   File: hd-disj-pressure.eps
%   Label: fig:pres-hd-disj
%
\begin{figure}
\includegraphics[width=\GraphicsWidth]{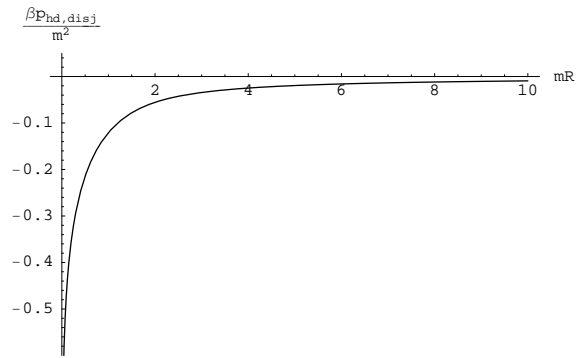}
\caption{
\label{fig:pres-hd-disj}
The disjoining pressure $p_{\text{hd,disj}}$ for the disk in the case
of non-attractive boundaries ($\alpha=0$) as a function of the radius
$R$. Notice that the disjoining pressure is always negative and it is
an increasing function of the radius, indicating that the system is
always unstable.
}
\end{figure}
%
%
%%%%%%%%%%%%%%%%%%%%%%%%%%%%%%%%%

A plot of the disjoining pressure $p_{\text{hd,disj}}$ for a hard
wall disk as a function of the radius $R$ is shown in
Figure~\ref{fig:pres-hd-disj}. Notice that $p_{\text{hd,disj}}$ is an
increasing function of $R$ and it is always negative. This shows that
in the absence of the attractive potential on the walls the systems is
always unstable for any radius $R$. This is a common feature of the
disk geometry with the slab geometry studied in our previous
work~\cite{TellezMerchan-jabon}. The system without any attractive
potential on the boundary ($\alpha=0$) is naturally unstable.

For $\alpha\neq0$, the disjoining pressure $p_d$ is given by
\begin{equation}
  p_d =  p_{\text{hd,disj}} + p_{\text{atd}}
\end{equation}
with $p_{\text{atd}}$ given by
equation~(\ref{eq:pressure-atd}). Before proceeding one should be
aware of an important fact. Although the first term
$p_{\text{hd,disj}}$ is finite when the cutoff $a\to0$ the second
term on the other hand is divergent when $a\to0$. The sum in
equation~(\ref{eq:pressure-atd}) should be cutoff to a upper limit
$N=R/a$ as it was done in the preceding section. Rigorously
speaking when $a\to 0$ the dominant term for the disjoining
pressure is $p_{\text{atd}}$. For large radius $R$, from last section
results~(\ref{eq:gd-pot-finite-size-exp-disk})
and~(\ref{eq:surface-tension-domin}) on the finite-size corrections we
can deduce the dominant term of the disjoining pressure when
$a\to 0$
\begin{eqnarray}\label{eq:pd-dominant-gamma}
  \beta p_d &\sim& -\frac{1}{R} \ \beta \gamma\nonumber\\
  &\sim& \frac{\alpha  m}{4\pi R}\ln \frac{2}{ma}
\end{eqnarray}
This term is always positive and is a decreasing function of $R$
indicating that the system is always stable.

This is an important difference between the slab geometry studied in
Ref.~\cite{TellezMerchan-jabon} and the present case of the disk. In
the slab geometry the disjoining pressure is always finite for
$a=0$ and any value of $\alpha$. For a slab of width $W$, the
large-$W$ expansion of the grand potential per unit area $\omega$
reads~\cite{TellezMerchan-jabon}
\begin{equation}
  \omega = -p_b W +  2\gamma + O\left(e^{-mW}\right)
\end{equation}
Then the pressure $p=\partial \omega/\partial W$ does not contain any
contribution from the surface tension. On the other hand in the disk
geometry considered here the existence of the curvature makes the
surface tension $\gamma$ very relevant for the disjoining pressure
(see equation~(\ref{eq:pd-dominant-gamma})) and since $\gamma$
diverges logarithmically with the cutoff it plays a dominant role in
the stability of the system.

Let us now consider a small but non-zero cutoff $a$. It is expected
that the results of our theory in that case should be close to the
ones of a model of small hard-core particles of diameter $a$. Then
there will be a competition between the natural unstable behavior of
the case without attractive potential ($\alpha=0$) and the attractive
part to the pressure $p_{\text{atd}}$ which is
stabilizing. 

\begin{figure}
\includegraphics[width=\GraphicsWidth]{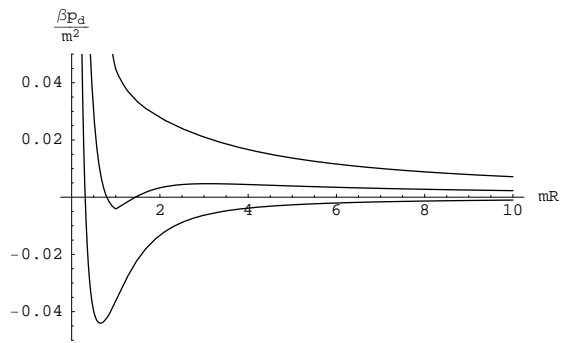}
\caption{
\label{fig:disj-pres-disk}
The disjoining pressure $p_d$ for the disk as a function of the radius
$R$ for $\alpha=0.15, 0.21$ and~$0.3$ from bottom to top. The cutoff is
chosen as $ma=10^{-3}$. The case $\alpha=0.3$ shows a stable disk
for all radius, in the case $\alpha=0.15$ only small disks are stable
and when $\alpha=0.21$ there is the possibility of a collapse from large
disks to small disks.  }
\end{figure}

Figure~\ref{fig:disj-pres-disk} shows several plots of the disjoining
pressure $p_d$ as a function of the radius $R$ for three special
values of the adhesivity $\alpha=0.15, 0.21$ and~$0.3$ for a fixed
cutoff $ma=10^{-3}$. These three plots show three characteristic
regimes in which the system can be.

For large values of the adhesivity, for example the case $\alpha=0.3$
shown in the Figure~\ref{fig:disj-pres-disk}, the disjoining pressure
is always positive and a decreasing function of the radius $R$. This
is a normal behavior: if the system is compressed the internal
pressure increases. This indicates a stable system for all values of
the radius $R$.

%%%%%%%%%%%%%%%%%%%%%%%%%%%%%%%%%%%%%%%%%%%%%%%%%%%%%%%%%%
%
%    Figure 2
%    File: dis-pressure.eps
%    Label: fig:disj-pres-disk
%
%
%
%%%%%%%%%%%%%%%%%%%%%%%%%%%%%%%%%%%%%%%%%%%%%%%%%%%%%%%%%%%%%%%%%

\begin{figure}
\includegraphics[width=\GraphicsWidth]{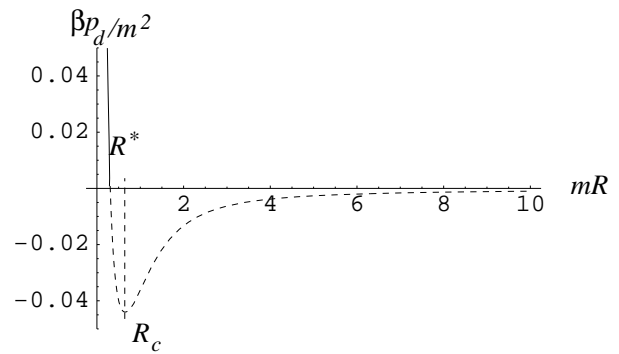}
\caption{
\label{fig:marginally-stable}
The disjoining pressure $p_d$ for the disk as a function of the radius
$R$ for $\alpha=0.15$. The cutoff is chosen as $ma=10^{-3}$.
The radius $R_c$ is defined by $\frac{\partial p_d}{\partial
  R}=0$. A very large disk $R\to\infty$ would collapse to a disk of
radius $R^*$ with disjoining pressure $p_d(R^*)=0$. The dashed region for
$R>R^*$ is not physical.
}
\end{figure}

In the case $\alpha=0.15$ (small values of the adhesivity) we have two
different behaviors of the disjoining pressure. For radius $R$ smaller
than a certain ``critical'' radius $R_c$, the disjoining pressure is a
decreasing function of the radius $R$, indicating again a stable
system. However, for radius $R$ larger than $R_c$ the disjoining
pressure is an increasing function of the radius $R$. This behavior is
anomalous, indicating that in this range of radius the system is not
stable. A very large disk $R\to\infty$ is marginally stable and will
collapse to a disk of smaller radius $R^*$ where $R^*$ is the radius
corresponding to $p_{d}=0$ (see Figure~\ref{fig:marginally-stable}).

%%%%%%%%%%%%%%%%%%%%%%%%%%%%%%%%%%%%%%%%%%%%%%%%%%%%%%%%%%%%%%%%%%%%%%%%
%
%    Figure 3
%    File: marginally-stable.eps
%    Label: fig:marginally-stable
%
%
%
%%%%%%%%%%%%%%%%%%%%%%%%%%%%%%%%%%%%%%%%%%%%%%%%%%%%%%%%%%%%

\begin{figure}
\includegraphics[width=\GraphicsWidth]{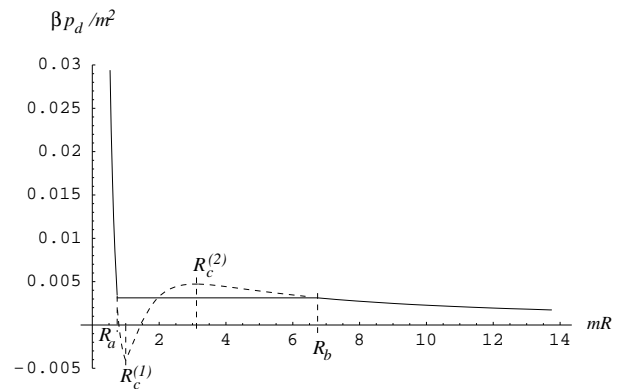}
\caption{
\label{fig:Maxwell-constr}
The disjoining pressure $p_d$ as a function of the radius $R$ for
$\alpha=0.21$ and a cutoff $ma=10^{-3}$. The theoretical result
for the pressure shows a non-physical region (dashed line). The
correct $p_d$ vs.~$R$ curve can be obtained by a Maxwell construction,
thus showing that in this case there is a first order transition
(collapse) from a large disk (radius $R_b$) to a small disk (radius
$R_a$).
}
\end{figure}

For $\alpha=0.21$ (intermediate values of the adhesivity) we have a
crossover regime. There are now three different behaviors of the
pressure characterized by two special radius $R_c^{(1)}$ and
$R_c^{(2)}$.  For $R<R_c^{(1)}$ the pressure is a decreasing function
of $R$: a stable regime. Then for $R_c^{(1)}<R<R_c^{(2)}$ the pressure
is an increasing function of $R$: an unstable regime. Finally for
$R>R_c^{(2)}$ the pressure is again a decreasing function of $R$:
another stable regime. One can do a Maxwell construction do determine
the correct dependency of the pressure against the radius as shown in
Figure~\ref{fig:Maxwell-constr}. In this regime there is a first order
transition (collapse) from a large (but finite) disk (radius $R_b$) to
smaller disk with radius $R_a$ (see Figure~\ref{fig:Maxwell-constr}).

%%%%%%%%%%%%%%%%%%%%%%%%%%%%%%%%%%%%%%%%%%%%%%%%%%%%%%%%%%%%
%
%
%    Figure 4
%    File: maxwell-construction.eps
%    Label: fig:Maxwell-constr
%
%
%
%%%%%%%%%%%%%%%%%%%%%%%%%%%%%%%%%%%%%%%%%%%%%%%%%%%%%%%%%%%%

Two of the three regimes illustrated here, small adhesivity (for
example $\alpha=0.15$) and large adhesivity (for example
$\alpha=0.3$), also occur for the slab geometry studied in
Ref.~\cite{TellezMerchan-jabon}. In that case they were separated by
the special value $\alpha=\alpha_c=1$. The disk case considered here
is however more rich since there is also a crossover regime between
the two, for intermediate values of $\alpha$ (for example
$\alpha=0.21$), with the possibility of stable large radius $R$ disks,
a forbidden (unstable) intermediate range of radius $R$ and then again
stable small disks.

The values of $\alpha$ characterizing the different regimes are
however highly dependent on the cutoff and so are the values of the
``critical'' radius $R_c$, $R_c^{(1)}$ and $R_c^{(2)}$. The values
$\alpha=0.15, 0.21$ and~$0.3$ of Figure~\ref{fig:disj-pres-disk} are
for a cutoff $ma=10^{-3}$, for other values of the cutoff these values
will change. For this reason the analysis above should be considered
with some care. For very small values of the cutoff $a$ we expect that
the results obtained here should reproduce asymptotically those of a
primitive model of hard-core particles with radius $a$.

To conclude this section let us mention that in the annulus geometry
the situation will be similar to the disk one. In that case one can
consider the pressure on the inner boundary or the pressure on the
outer boundary. For both pressures the dominant term when the cutoff
$a\to 0$ will be given in terms of the surface tension $\gamma$
as in equation~(\ref{eq:pd-dominant-gamma}). Then in the strict limit
$a\to0$ the annulus geometry will be stable if $\alpha>0$ as in
the disk case. Considering a small but non-zero cutoff $a$ will
lead to a similar discussion as in the disk with different regimes
some of them exhibiting a collapse.

%%%%%%%%%%%%%%%%%%%%%%%%%%%%%%%%%%%%%%%%%%%%%%%%%%%%%%%%%%%%%%%

\section{The density profiles}
\label{sec:Density}

In this section, we study the densities of the Coulomb system confined
inside a disk and inside an annulus.  The densities can be obtained by
computing the Green functions introduced in
Sec.~\ref{sec:Model}. First we will solve and extensively study model
I for the disk then the annulus. Then we will briefly consider the
results that can be obtained using model II.
 
\subsection{The disk}

In the disk geometry, for model I, the fugacity $m_{-}(\r)$ is given
by Eq.~(\ref{eq:m-disk}). From Eq.~(\ref{eq:GreenFunctions-G}), when
$r_2\not=R$, we see that $G_{--}$ and $G_{-+}$ are continuous for
$r_1=R$. However because of the Dirac delta distribution in the
definition of $m_-(r)$ the functions $G_{+-}$ and $G_{++}$ are
discontinuous at $r_1=R$. The discontinuity can be obtained from
Eq.~(\ref{eq:GreenFunctions-G}) if $r_2\neq r_1$:
\begin{eqnarray}
\label{eq:disc}
G_{+\mp}(r_1=R^-,r_2)&-& G_{+\mp}(r_1=R^+,r_2)\nonumber\\
&=&\alpha~ e^{-i\theta_1}G_{-\mp}(r_1=R,r_2)\nonumber\\
\end{eqnarray}

If both points $\r_1$ and $\r_2$ are inside the disk but not on the
boundary then Eq.~(\ref{eq:GreenFunctions-G}) lead to a Helmoltz
equation for $G_{++}$ and for $G_{--}$
\begin{equation}
\label{eq:delta-G}
\left[ m^{2}-\Delta \right] 
G_{\pm \pm }\left( \r_{1},\r_{2}\right) 
= m\delta \left( \r_{1}-\r_{2}\right)
\end{equation}
and the other Green functions can be obtained from
\begin{equation}
\label{eq:homogenea-G}
\frac{e^{\pm i\theta_1}}{m}
\left(-\partial_{r_1}\mp\frac{i}{r_1}\partial_{\theta_1}\right)G_{\pm
\pm }\left( \r_{1},\r_{2}\right) = G_{\mp \pm }\left(
\r_{1},\r_{2}\right)
\end{equation}
Equation~(\ref{eq:delta-G}) has solutions of the form
\begin{eqnarray}
\label{eq:form-solut-disk}
G_{\pm\pm}(\r_1,\r_2)&=& \frac{m}{2\pi} \sum_{l=-\infty}^{+\infty}
e^{il(\theta_1-\theta_2)}\left[ I_l(mr_<)K_l(mr_>)\right.\nonumber\\
&+&\left. A^{\pm}_l(r_2) I_l(mr_1)\right]\nonumber\\
\end{eqnarray}
where $r_< =\min(r_1,r_2)$ and $r_> =\max(r_1,r_2)$.  The remaining
Green functions can be found using Eq.~(\ref{eq:homogenea-G}).

If $\r_1$ is outside the film while $\r_2$ is fixed inside the film
$m_+(r_1)=m_-(r_1)=0$, the $G_{s_1s_2}$ that satisfy
Eq.~(\ref{eq:GreenFunctions-G}) are
\begin{eqnarray}
G_{+\mp}(\r_1,\r_2)&=&\sum_{l=-\infty}^{+\infty} C_l(r_2,\theta_2)(
r_1e^{i\theta_1})^l\,,\\
G_{-\mp}(\r_1,\r_2)&=&\sum_{l=-\infty}^{+\infty} D_l(r_2,\theta_2) (
r_1e^{-i\theta_1})^{-l}\,.
\end{eqnarray}
So, in order to have finite solutions at $r_1=\infty$ it is necessary
that for $r_1>R$
\begin{equation}
\label{eq:fuera}
\begin{array}{cc}
C_l(r_2,\theta_2)=0 & \text{ for $l\geq0$}\\
D_l(r_2,\theta_2)=0 & \text{ for $l\leq0$}\,,\\
\end{array}
\end{equation}
Eqs.~(\ref{eq:disc}) and~(\ref{eq:fuera}) are the boundary conditions
that complement the differential equations~(\ref{eq:delta-G})
and~(\ref{eq:homogenea-G}) for the Green functions.

Solving for the coefficients $A_{l}^{\pm}$, we arrive at the following
expressions for the Green functions for $0 \leq r_{1,2}< R$:
\begin{eqnarray}
\label{eq:result-G++-disk}
G_{++}(\r_1,\r_2)&=&\frac{m}{2\pi}K_0(m|\r_1-\r_2|)\nonumber\\
&+&\frac{m}{2\pi}\sum_{l=0}^{+\infty}e^{il(\theta_1-\theta_2)}
\Bigg[
\frac{K_l}{I_l}
 I_{l+1}(mr_1) I_{l+1}(mr_2)\nonumber\\
&+&
\frac{\alpha K_{l+1}-K_l}{\alpha I_{l+1}+I_l}I_{l}(mr_1)I_{l}(mr_2)
\Bigg]
\end{eqnarray}
and
\begin{eqnarray}
\label{eq:result-G---disk}
G_{--}(\r_1,\r_2)&=&\frac{m}{2\pi}K_0(m|\r_1-\r_2|) \nonumber\\
&-&\frac{m}{2\pi}\sum_{l=0}^{+\infty}e^{il(\theta_1-\theta_2)}
\Bigg[
\frac{K_l}{I_l}
 I_{l}(mr_1) I_{l}(mr_2)\nonumber\\
&+&
\frac{\alpha K_{l+1}-K_l}{\alpha I_{l+1}+I_l}I_{l+1}(mr_1)I_{l+1}(mr_2)
\Bigg]
\end{eqnarray}
with $I_{l}$ meaning $I_{l}(mR)$, and the same convention for the
Bessel function $K_l$.  

The one-particle densities are given in terms of these Green functions as 
\begin{equation}
\rho_{s}(\r)=m_{s}(r)G_{s s}(\r,\r)
\end{equation}
As we explained in section~\ref{sec:Model} the continuous limit model
presents divergences in the expressions of the densities. This is seen
in the term $K_0(mr_{12})$ that diverges logarithmically as
$r_{12}=|\r_1-\r_2|\rightarrow 0$. So we impose a short distance
cutoff $a$. One can think that particles are disks of diameter $a$, so
the minimal distance between particles is $a$.

The first term in Eqs.~(\ref{eq:result-G++-disk})
and~(\ref{eq:result-G---disk}) gives the bulk density $\rho_b$, the
density of the unbounded system as calculated in
Ref.~\cite{CornuJanco}. For $a\rightarrow 0$
\begin{equation}
\label{eq:bulk-density}
\rho_{b}^{+}=\rho_{b}^{-}
=
\rho_{b}=\frac{m^2}{2\pi}K_0(ma) \sim
\frac{m^2}{2\pi}\left[\ln\frac{2}{ma}-\gamma\right]
\end{equation}
where $\gamma\simeq0.5772$ is the Euler constant.  In the second terms
of Eqs.~(\ref{eq:result-G++-disk}) and~(\ref{eq:result-G---disk}) the
sum can eventually diverge when $\r_1=\r_2=\r$ for certain values of
$\r$ (in the boundaries) so we should impose a cutoff $|l|<N=R/a$ as
it has been done in the expressions of the pressure and
grand-potential obtained in the last section.

Because of the form~(\ref{eq:m-disk}) of $m_{-}(r)$, the negative
density can be written as
\begin{equation}
  \rho_{-}(r)=\left(1+\frac{\alpha}{m}\, \delta(r-R)\right) \rho^{*}_{-}(r)
\end{equation}
where $\rho^{*}_{-}$ can be seen as the density of non-adsorbed
particles. For the positive particles $\rho_{+}=\rho_{+}^{*}$. Finally
we have
\begin{widetext}
\begin{subequations}
\begin{eqnarray}
\label{eq:rho+result-final}
  \rho_{+}^{*}(r)&=&\rho_{b}+\frac{m^2}{2\pi}
\sum_{l=0}^{\infty}
\left[
  \frac{K_l}{I_l}\, I_{l+1}^2(mr)+
  \frac{\alpha K_{l+1}-K_l}{\alpha I_{l+1}+I_l}\,I_{l}^2(mr)
\right]
\\
\label{eq:rho-result-final}
  \rho_{-}^{*}(r)&=&\rho_{b}-\frac{m^2}{2\pi}
\sum_{l=0}^{\infty}
\left[
  \frac{K_l}{I_l}\, I_{l}^2(mr)+
  \frac{\alpha K_{l+1}-K_l}{\alpha I_{l+1}+I_l}\,I_{l+1}^2(mr)
\right]
\end{eqnarray}
\end{subequations}
\end{widetext}
The non-adsorbed charge density $\rho^{*}=\rho_{+}^*-\rho_{-}^{*}$ can
be obtained from the above expression and using the Wronskian of the
Bessel functions $I_l K_{l+1}+I_{l+1} K_{l}=1/mR$,
\begin{equation}
  \rho^{*}(r)= \frac{\alpha m^2}{2\pi R} \sum_{l=0}^{\infty}
  \frac{I_{l+1}^2(mr)+I_{l}^2(mr)}{(\alpha I_{l+1}+I_l)I_l}
\end{equation}
Finally, the total charge density
\begin{eqnarray}
\rho(r)&=&\rho_{+}(r)-\rho_{-}(r)\nonumber\\ 
&=&\rho_{+}(r)-
\left(1+\frac{\alpha}{m}\,\delta(r-R)\right)\rho^{*}_{-}(r)\\
&=&
\rho^{*}(r)-\sigma_{-} \delta(r-R)
\nonumber
\end{eqnarray}
has a non-adsorbed part $\rho^{*}(r)$ and a adsorbed ``surface''
charge density in the boundary
\begin{equation}
\label{eq:adsorbed-charge-1}
\sigma_{-}=\frac{\alpha}{m}\,\rho^{*}_{-} (R)  
\end{equation}
This surface charge density $\sigma_{-}$ comes from the $\delta(r-R)$
part of the negative charge density. Writing formally the bulk part of
the density as $\rho_b=(m^2/2\pi) \sum_{l\in\mathbb{Z}} I_l K_l$ one
can obtain the following expression for $\sigma_{-}$ from
Eqs.~(\ref{eq:adsorbed-charge-1}) and~(\ref{eq:rho-result-final})
\begin{equation}
\label{eq:adsorbed-surface-charge-result}
  \sigma_{-}=\frac{\alpha}{2\pi R}
  \sum_{l=0}^{\infty} \frac{I_{l+1}}{\alpha I_{l+1} + I_{l}}
\end{equation}

Actually the adsorbed charge density should obey two special
relations. The first is a sum rule that expresses the global
electro-neutrality of the system
\begin{equation}
\label{eq:sum-rule-electroneutrality}
  R\, \sigma_{-}=\int_0^{R} \rho^{*}(r)\,r\,dr
\end{equation}
Using the indefinite integral
\begin{widetext}
\begin{equation}
  \label{eq:int-II}
  \int^{mR} x (I_{l+1}^2(x)+I_{l}^2(x))\,dx
  =\frac{(mR)^2}{2} \left[ I_{l+1}^2-I_{l} I_{l+2}+
  I_{l}^2-I_{l-1} I_{l+1}\right] 
  = 
  mR\, I_{l} I_{l+1}  
\end{equation}
\end{widetext}
obtained using the recurrence relations~\cite{Grad} for the Bessel
functions $I_{l}$, the sum rule~(\ref{eq:sum-rule-electroneutrality})
is immediately shown to be satisfied.

On the other hand the adhesivity $\alpha$ can be thought as a sort of
fugacity that controls the number of adsorbed
particles~\cite{RosinbergLebowitzBlum} and one can obtain the total
number of adsorbed particles $2\pi R \sigma_{-}$ from the grand
potential $\Omega$ by using the usual thermodynamic relation
\begin{equation}
  \label{eq:sigma-from-Omega-disk}
  2\pi R\,
  \sigma_{-}=-\alpha\beta\frac{\partial\Omega}{\partial
  \alpha}
\end{equation}
This relationship is also immediately shown to be satisfied from the
expression~(\ref{eq:grand-pot-disk-attract}) for
$\Omega_{\text{at}}^{\text{D}}$, the part of the grand potential that
depends on $\alpha$.

For a large disk, $R\to\infty$, the dominant part of the grand potential
that depends on $\alpha$ is the surface tension $\gamma$ given by
equations~(\ref{eq:surface-tension})
and~(\ref{eq:surface-tension-domin}). Then we have
\begin{equation}
  \label{eq:sigma-surface-tension}
  \sigma_{-}=-
  \beta\alpha\frac{\partial\gamma}{\partial\alpha}
\end{equation}
a relation already shown to be true in Ref.~\cite{TellezMerchan-jabon}
for the same system near a plane attractive hard wall. Using
Eq.~(\ref{eq:surface-tension}) into
Eq.~(\ref{eq:sigma-surface-tension}) gives explicitly 
\begin{equation}
  \sigma_{-}=\frac{m}{4\pi}\left[\alpha \ln\frac{2}{ma} -
  \frac{\alpha^2+1}{\alpha} \ln(\alpha+1)+1\right]
\end{equation}
Thus recovering a known result from Ref.~\cite{TellezMerchan-jabon}.
Notice that for large disks the adsorbed surface charge density
becomes independent of the radius $R$. This result can of course by
also obtained directly from
Eq.~(\ref{eq:adsorbed-surface-charge-result}) in the limit of
large-$R$ using the Debye expansions~(\ref{eq:Debye-exp-BesselI}) of
the Bessel functions.

%%%%%%%%%%%%%%%%%%%%%%%%%%%%%%%%%%%%%%%%%%%%%%%%%%%%%%%%%%%%
%
%
%    Figure 5
%    File: charge-density-disk-R10.eps
%    Label: fig:charge-density-disk-R10
%
\begin{figure}
\includegraphics[width=\GraphicsWidth]{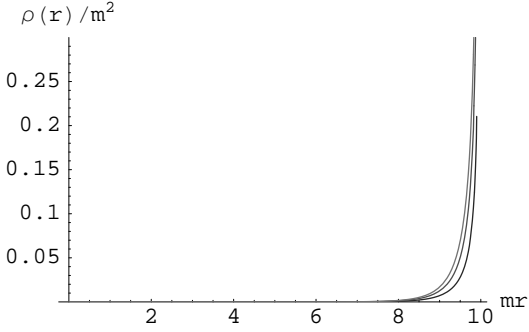}
\caption{
\label{fig:charge-density-disk-R10}
The charge density profile $\rho(r)$ for a disk of radius
$R=10/m$. From bottom to top the adhesivity $\alpha=0.25,$ 0.5, 0.75.
}
\end{figure}
%
%
%%%%%%%%%%%%%%%%%%%%%%%%%%%%%%%%%%%%%%%%%%%%%%%%%%%%%%%%%%%%

%%%%%%%%%%%%%%%%%%%%%%%%%%%%%%%%%%%%%%%%%%%%%%%%%%%%%%%%%%%%
%
%
%    Figure 6
%    File: densities-disk-R10.eps
%    Label: fig:densities-disk-R10
%
\begin{figure}
\includegraphics[width=\GraphicsWidth]{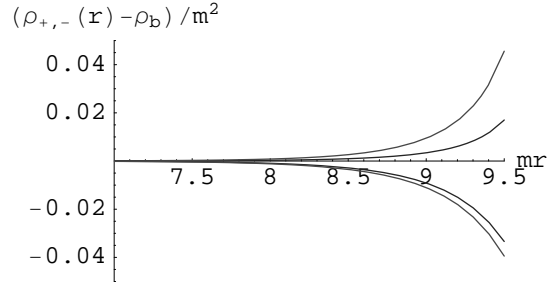}
\caption{
\label{fig:densities-disk-R10}
The density profiles $\rho_+(r)$ and $\rho_{-}(r)$ for a disk of
radius $R=10/m$. The bulk density has been subtracted from both
densities. The upper curves correspond to the density of the positive
particles $\rho_+(r)$ and the bottom curves to the density of the
negative particles $\rho_-(r)$.  The adhesivity $\alpha=1$ for the
topmost curve ($\rho_{+}(r)$) and the bottommost curve
($\rho_{-}(r)$). The two other curves correspond to $\alpha=0.5$.  }
\end{figure}
%
%
%%%%%%%%%%%%%%%%%%%%%%%%%%%%%%%%%%%%%%%%%%%%%%%%%%%%%%%%%%%%

In Figure~\ref{fig:charge-density-disk-R10}, we can see the charge
density profile $\rho(r)$ (actually only the non-adsorbed part
$\rho^{*}(r)$ is shown). It can be seen that the negative surface
charge $-\sigma_{-}$ (not shown in the figure) is screened by a
positive layer near $r=R$. As the adhesivity is increased,
$\sigma_{-}$ increases, and therefore the density of the screening
positive layer shown in Figure~\ref{fig:charge-density-disk-R10}
increases. This screening layer has a thickness of order of magnitude
$m^{-1}$ as expected since $m^{-1}$ is a measure of the screening
length~\cite{CornuJanco}.

In Figure \ref{fig:densities-disk-R10}, we can see the distribution of
the negative particles (two lower curves) and positive particles (two
upper curves). The negative particles feel repelled by the ones that
lie in the boundary (the surface charge $\sigma_{-}$). Therefore, the
negative density $\rho_{-}(r)$ decreases near the frontier. The
positive charge density $\rho_{+}(r)$ inside the disk can be observed
in the upper curves of Figure~\ref{fig:densities-disk-R10}. Due to
the large accumulation of negative charges at the frontier, the
positive particles tend to shield this charge. Hence, there is an
increase of the positive density near $R$.

As we increase the adhesivity, the negative adsorbed charge on the
border gets larger while on the inner region there is a stronger
repulsion of the negative charges and an increase in the number of
positive particles shielding the outer charge.

%%%%%%%%%%%%%%%%%%%%%%%%%%%%%%%%%%%%%%%%%%%%%%%%%%%%%%%%%%%%%%%

\subsection{The Annulus}

For the annulus geometry with a delta distribution modeling the
external attractive potential (model I), we follow the same reasonings
as for the disk geometry. The solution to Eq.~(\ref{eq:delta-G}) for
$R_1<r<R_2$ is of the form
\begin{eqnarray}
\label{eq:form-solut-annulus}
G_{\pm\pm}(\r_1,\r_2)&=& \frac{m}{2\pi} \sum_{l=-\infty}^{+\infty}
e^{il\theta_{12}}\left[ I_l(mr_<)K_l(mr_>)\right.\nonumber\\
&+&\left. A^{\pm}_l(r_2) I_l(mr_1)
+B^{\pm}_l(r_2) K_l(mr_1)
\right]\nonumber\\
\end{eqnarray}
where we defined $\theta_{12}=\theta_1-\theta_2$.
As in the disk case, the Green functions $G_{--}$
and $G_{-+}$ are continuous while $G_{++}$ and $G_{+-}$ are
discontinuous at $r=R_1$ and $r=R_2$ with a discontinuity given by
Eq.~(\ref{eq:disc}) replacing the adhesivity $\alpha$ by $\alpha_1$
for the discontinuity at $r=R_1$ and by $\alpha_2$ for the
discontinuity at $r=R_2$.

Solving for the coefficients we finally find
\begin{widetext} 
\begin{subequations}
\label{eq:result-G--annulus}
\begin{eqnarray}
  G_{--}(\r_1,\r_2)&=& \frac{m}{2\pi}
    K_0(m|\r_1-\r_2|)\\
    &-& \frac{m}{2\pi}
    \sum_{l=0}^{\infty} 
    \frac{e^{-il\theta_{12}}
	A_{l}^{(1)}
    }{D_{l}^{(1)}}
    I_l(mr_1) I_l(mr_2)\\
    &-&
    \frac{m}{2\pi}
    \sum_{l=0}^{\infty} 
    \frac{e^{-il\theta_{12}}
      B_{l}^{(1)}
    }{D_{l}^{(1)}}
    K_l(mr_1) K_l(mr_2)\\
    &+&
    \frac{m}{2\pi}
    \sum_{l=0}^{\infty}
    \frac{e^{-il\theta_{12}}
      C_{l}^{(1)}
    }{D_{l}^{(1)}}      
    \left[I_l(mr_1) K_l(mr_2)+K_l(mr_1) I_l(mr_2)\right]\\
    &+&
    \frac{m}{2\pi}
    \sum_{l=0}^{\infty}
    \frac{e^{i(l+1)\theta_{12}}
	A_{l}^{(2)}
      }{D_{l}^{(2)}}
     I_{l+1}(mr_1) I_{l+1}(mr_2)\\
    &+&
    \frac{m}{2\pi}
    \sum_{l=0}^{\infty}
    \frac{e^{i(l+1)\theta_{12}}
      B_{l}^{(2)} 
    }{D_{l}^{(2)}}
    K_{l+1}(mr_1) K_{l+1}(mr_2)\\
    &-&
    \frac{m}{2\pi}
    \sum_{l=0}^{\infty}
    \frac{e^{i(l+1)\theta_{12}}
      C_{l}^{(2)}      
    }{D_{l}^{(2)}}
     \left[ K_{l+1}(mr_1) I_{l+1}(mr_2)  + I_{l+1}(mr_1)
    K_{l+1}(mr_2)\right]
    \nonumber\\
\end{eqnarray}
\end{subequations}
and 
\begin{subequations}
\label{eq:result-G++annulus}
\begin{eqnarray}
  G_{++}(\r_1,\r_2)&=& \frac{m}{2\pi}
    K_0(m|\r_1-\r_2|)\\
    &+& \frac{m}{2\pi}
    \sum_{l=0}^{\infty} 
    \frac{e^{-i(l+1)\theta_{12}}
      A_{l}^{(1)}
    }{ D_{l}^{(1)}
    } 
    I_{l+1}(mr_1) I_{l+1}(mr_2) \\
    &+& \frac{m}{2\pi}
    \sum_{l=0}^{\infty} 
    \frac{e^{-i(l+1)\theta_{12}}
	B_{l}^{(1)}
    }{
	D_l^{(1)}
    } 
    K_{l+1}(mr_1) K_{l+1}(mr_2) \\
    &+& \frac{m}{2\pi}
    \sum_{l=0}^{\infty} 
    \frac{e^{-i(l+1)\theta_{12}}
	C_{l}^{(1)}
    }{
	D_{l}^{(1)}
    }\left[ I_{l+1}(mr_1)K_{l+1}(mr_2) +  K_{l+1}(mr_1) I_{l+1}(mr_2)
      \right]
\nonumber\\&&
\\	
    &-& \frac{m}{2\pi}
    \sum_{l=0}^{\infty} 
    \frac{e^{il\theta_{12}}
      A_{l}^{(2)}
    }{
	D_{l}^{(2)}
    }
    I_{l}(mr_1) I_{l}(mr_2)\\
    &-& \frac{m}{2\pi}
    \sum_{l=0}^{\infty} 
    \frac{e^{il\theta_{12}}
	B_{l}^{(2)}
    }{
	D_{l}^{(2)}
    }
    K_{l}(mr_1)    K_{l}(mr_2) \\
    &-& \frac{m}{2\pi}
    \sum_{l=0}^{\infty} 
    \frac{e^{il\theta_{12}}
	C_{l}^{(2)}
    }{
	D_{l}^{(2)}
    }
      \left[
	I_l(mr_1) K_l(mr_2) + 	K_l(mr_1) I_l(mr_2) \right]
\end{eqnarray}
\end{subequations}
with
\begin{subequations}
\begin{eqnarray}
A_{l}^{(1)}&=&K_{l}^{(2)}\left( 
      \alpha_1 K_{l}^{(1)} + K_{l+1}^{(1)}  \right)\\
B_{l}^{(1)}&=& I_l^{(2)}\left(\alpha_1 I_{l}^{(1)} -
      I_{l+1}^{(1)}\right)\\
C_{l}^{(1)}&=& K_l^{(2)}
      \left( \alpha_1 I_l^{(1)} - I_{l+1}^{(1)} \right)\\
D_{l}^{(1)}&=&I_l^{(2)}\left( \alpha_1 K_l^{(1)} + K_{l+1}^{(1)}
      \right) - K_l^{(2)}\left( \alpha_1 I_l^{(1)} - I_{l+1}^{(1)}
      \right)\\
A_{l}^{(2)}&=& K_{l+1}^{(1)}
	\left(\alpha_2 K_{l+1}^{(2)}- K_{l}^{(2)}\right)\\
B_{l}^{(2)}& = & I_{l+1}^{(1)} 
      \left( \alpha_2 I_{l+1}^{(2)} + I_{l}^{(2)} \right )\\
C_{l}^{(2)} &=& I_{l+1}^{(1)}
\left(\alpha_2 K_{l+1}^{(2)} - K_{l}^{(2)} \right)\\
D_{l}^{(2)}&=&	I_{l+1}^{(1)}
	\left( \alpha_2 K_{l+1}^{(2)} - K_{l}^{(2)}\right)
	- K_{l+1}^{(1)}
	\left( \alpha_2 I_{l+1}^{(2)} + I_l^{(2)}\right)
\end{eqnarray}
\end{subequations}
\end{widetext}
where $I_{l}^{(1)}=I_{l}(mR_1)$, $I_{l}^{(2)}=I_{l}(mR_2)$ and the
same convention for the other Bessel functions.

The individual densities are obtained putting $\r_1=\r_2=\r$ in the
above expressions since $\rho_{s}(r)=m_s(\r) G_{ss}(\r,\r)$.  The
above expressions are quite long, however one can notice some
symmetries between $G_{--}$ and $G_{++}$. Beside a change of sign and
the phase factor (which is irrelevant in the calculation of the
densities) the coefficient of the term $I_l (mr_1)I_l (mr_2)$ in
$G_{--}$ is the same as the one for $I_{l+1}(mr_1)I_{l+1} (mr_2)$ in
$G_{++}$ and so on. The charge density can be written again as
\begin{equation}
  \rho(r)=\rho^{*} (r) - \sigma_{-}^{(1)} \delta(r-R_1)
  - \sigma_{-}^{(2)} \delta(r-R_2)
\end{equation}
with a non-aborbed part $\rho^{*}(r) $ and a surface density
\begin{equation}
\label{eq:surface-charge-annulus}
\sigma_{-}^{(1,2)}=(\alpha_{1,2}/m) \rho_{-}^{*}(R_{1,2})
\end{equation}
of adsorbed negative particles in $R_1$ and $R_2$ respectively.  The
non-adsorbed charge density is
\begin{widetext}
\begin{eqnarray}
  \rho^{*}(r)=\frac{m^2}{2\pi}
  \sum_{l=0}^{\infty} & \Bigg[ &
  (C_{1,l}^{II} + C_{2,l}^{II})\left[I_{l+1}^2(mr) + I_{l}^2(mr)\right] +
  (C_{1,l}^{KK} + C_{2,l}^{KK})\left[K_{l+1}^2(mr) +
  K_{l}^2(mr)\right]
    \nonumber\\ 
  && +  2 (C_{1,l}^{IK}+C_{2,l}^{IK})
  \left[I_{l+1}(mr) K_{l+1}(mr) - I_{l}(mr) K_{l}(mr) \right]\Bigg]
\end{eqnarray}
\end{widetext}
with the coefficients that depend on $\alpha_1$ given by
\begin{eqnarray}
  C_{1,l}^{II}=
  \frac{A_{l}^{(1)}}{D_{l}^{(1)}}\,,\
  &\displaystyle
  C_{1,l}^{KK}=
  \frac{B_{l}^{(1)}}{D_{l}^{(1)}}\,,\
	&
  C_{1,l}^{IK}=
  \frac{C_{l}^{(1)}}{D_{l}^{(1)}}\ \ \ 
\end{eqnarray}
and the ones that depend on $\alpha_2$ are
\begin{subequations}
\begin{eqnarray}
 C_{2,l}^{II}=
   -\frac{A_{l}^{(2)}}{D_{l}^{(2)}}
,
  &\displaystyle
   C_{2,l}^{KK}=
   -\frac{B_{l}^{(2)}}{D_{l}^{(2)}}
, &
   C_{2,l}^{IK}=
    \frac{C_{l}^{(2)}}{D_{l}^{(2)}}\ \ \ 
\end{eqnarray}
\end{subequations}
The adsorbed charge density in each boundary can be computed by
replacing $G_{--}(R_{1,2},R_{1,2})$ from
Eq.~(\ref{eq:result-G--annulus}) into
Eq.~(\ref{eq:surface-charge-annulus}) or by using the thermodynamic
relation
\begin{equation}
  \label{eq:sigma-from-Omega}
  2\pi R_{1,2}\,
  \sigma_{-}^{(1,2)}=-\alpha_{1,2}\beta
  \frac{\partial\Omega^{\text{A}}}{\partial
  \alpha_{1,2}}
\end{equation}
Either way the result is the same as expected
\begin{subequations}
\label{eq:sigmas}
\begin{eqnarray}
  \label{eq:sigma1}
  \sigma_{-}^{(1)}
  &=& \frac{\alpha_1}{2\pi R_1}
  \sum_{l=0}^{\infty}
  \frac{I_l^{(2)} K_l^{(1)}-K_l^{(2)} I_l^{(1)}
  }{D_{l}^{(1)}}
\\
  \label{eq:sigma2}
  \sigma_{-}^{(2)}
  &=& \frac{\alpha_2}{2\pi R_2}
  \sum_{l=0}^{\infty}
  \frac{I_{l+1}^{(2)} K_{l+1}^{(1)} - K_{l+1}^{(2)} I_{l+1}^{(1)}
  }{D_{l}^{(2)}}
\end{eqnarray}
\end{subequations}
Using the indefinite integrals~(\ref{eq:int-II}) and~\cite{Watson}
\begin{widetext}
\begin{eqnarray}
  \int^{mR} (K_{l}^2(x) 
  + K_{l+1}^2(x))\,x\,dx &=&-mR\, K_{l+1} K_{l}\\
  \int^{mR} (I_{l+1}(x) K_{l+1}(x) - I_{l}(x) K_{l} (x) )\,x\,dx
  &=& \frac{mR}{2}(K_{l+1} I_{l}-I_{l+1} K_{l})
\end{eqnarray}
\end{widetext}
one can verify that the electroneutrality sum rule
\begin{equation}
  \label{eq:electroneutrality-sum-rule-annulus}
  \int_{R_1}^{R_2} \rho^{*}(r)\, r\, dr= R_1\, \sigma_{-}^{(1)} + 
  R_2\, \sigma_{-}^{(2)}
\end{equation}
is satisfied. For the reader interested in the details of this
calculation, an extented version of this manuscript is available
online~\cite{Merchan-Tellez-tcp-anillos-v1}.

%%%%%%%%%%%%%%%%%%%%%%%%%%%%%%%%%%%%%%%%%%%%%%%%%%%%%%%%%%%%
%
%
%    Figure 6
%    File: rhoannulus.eps
%    Label: fig:rhoannulus
%
\begin{figure}
\includegraphics[width=\GraphicsWidth]{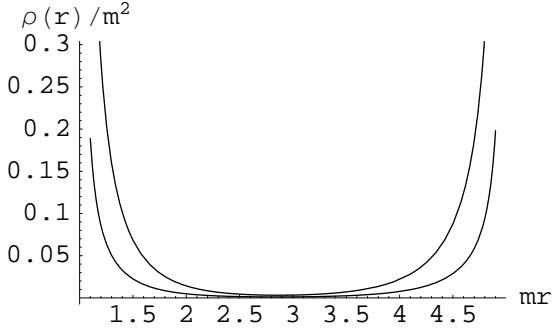}
\caption{
\label{fig:rhoannulus}
The charge density profile $\rho(r)$ for an disk of inner radius
$R_1=1/m$ and outer radius $R_2=5/m$. The adhesivities in each
boundary have been chosen equals $\alpha_1=\alpha_2=\alpha$.  The
upper curve correspond to $\alpha=1$ while the lower one to
$\alpha=0.25$.  }
\end{figure}
%
%
%%%%%%%%%%%%%%%%%%%%%%%%%%%%%%%%%%%%%%%%%%%%%%%%%%%%%%%%%%%%

\begin{figure}
\includegraphics[width=\GraphicsWidth]{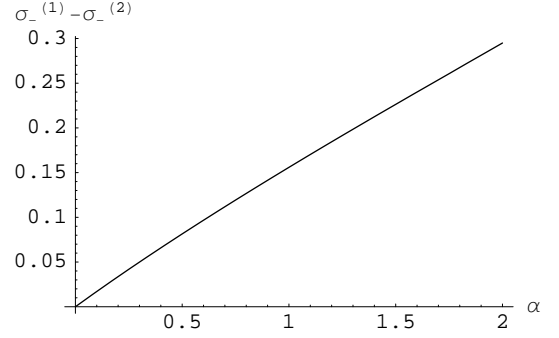}
\caption{
\label{fig:diffsigmas}
The difference between the surface charge in the inner boundary and
the surface charge in the outer one
$\sigma_{-}^{(1)}-\sigma_{-}^{(2)}$ as a function of $\alpha$ for an
annulus of inner radius $R_1=1/m$ and outer radius $R_2=5/m$.
}
\end{figure}

Figure~\ref{fig:rhoannulus} shows a plot of the charge density
$\rho(r)$ for an annulus with inner radius $R_1=1/m$ and outer radius
$R_2=5/m$ for $\alpha_1=\alpha_2=0.5$ and $\alpha_1=\alpha_2=1$. In
both figures on can see a positive layer of charge in each boundary
screening the adsorbed negative surface charge densities
$\sigma_{-}^{(1,2)}$. As $\alpha_{1,2}$ increases the adsorbed charge
increases and so does the positive layer. Although it is not perfectly
clear in Figure~\ref{fig:rhoannulus} there is actually slightly more
adsorbed surface charge density in the inner boundary that in the
outer. This can be seen in Figure~\ref{fig:diffsigmas} that shows the
difference between the adsorbed surface charge in the inner boundary
and the surface charge in the outer one,
$\sigma_{-}^{(1)}-\sigma_{-}^{(2)}$, as a function of
$\alpha$. Figure~\ref{fig:diffsigmas} clearly shows that
$\sigma_{-}^{(1)}>\sigma_{-}^{(2)}$ if $\alpha_{1}=\alpha_{2}$. On the
other hand the total charge on the inner boundary is smaller that the
total charge on the outer boundary: $2\pi R_{1}\, \sigma_{-}^{(1)} <
2\pi R_{2}\, \sigma_{-}^{(2)}$. This can be seen directly from
Eqs.~(\ref{eq:sigmas}). If $\alpha_1=\alpha_2=\alpha$, each term in
the series of the difference $R_1\, \sigma_{-}^{(1)}- R_2\,
\sigma_{-}^{(2)}$ from Eqs.~(\ref{eq:sigmas}) is of the form
\begin{equation}
  \frac{\alpha}{D_l^{(1)} D_l^{(2)}}\,\left( b_{l}-b_{l+1}\right)
\end{equation}
where $D_{l}^{(1)}$ and $D_{l}^{(2)}$ are the denominators in each
term of the sums in Eqs.~(\ref{eq:sigma1}) and~(\ref{eq:sigma2})
respectively, and $b_{l}=K_{l}^{(1)} I_{l}^{(2)}-K_{l}^{(2)}
I_{l}^{(1)}$. The sequence $(b_{l})_{l\in\mathbb{N}}$ has the property
of being monotonically increasing with $l$. Then we conclude $R_1\,
\sigma_{-}^{(1)} < R_2\, \sigma_{-}^{(2)}$.

%%%%%%%%%%%%%%%%%%%%%%%%%%%%%%%%%%%%%%%%%%%%%%%%%%%%%%%%%%%%
%
%
%    Figure 7
%    File: diffsigmas.eps
%    Label: fig:diffsigmas
%
%
%
%%%%%%%%%%%%%%%%%%%%%%%%%%%%%%%%%%%%%%%%%%%%%%%%%%%%%%%%%%%%

%%%%%%%%%%%%%%%%%%%%%%%%%%%%%%%%%%%%%%%%%%%%%%%%%%%%%%%%%%%%

\subsection{Model II}
\label{sec:modelII}

We now briefly consider model II and some of its results in the
annulus geometry. 

In the annulus geometry, there are two outer regions, each one has a
width $\deltaw$, and an inner region that has a thickness
$R_2-R_1-2\deltaw$. We can distinguish three regions for the annulus
geometry: the inner border $R_1<r_1<R_1+\deltaw$ (region 1), the bulk
of the film $R_1+\deltaw<r_1<R_2-\deltaw$ (region 2) and the outer
border $R_2-\deltaw<r_1<R_2$ (region 3).  So for this model in the
annulus geometry, we define the position dependent fugacities as
\begin{subequations}
\label{eq:m-annulus-step}
\begin{eqnarray}
m_+(r_1)&=&m\\
m_-(r_1)&=&
\begin{cases}
m_2 & \text{if $r_1\in$ region 1}\\
m & \text{if $r_1 \in$ region 2}\\
m_2   & \text{if $ r_1 \in$ region 3}\\
\end{cases}
\end{eqnarray}
  
\end{subequations}

The fugacity in the border regions is $m_2=m\exp(-\beta U_{-})$ where
$U_{-}<0$ is the value of the external potential $V_{-}(\r)$ near the
boundary. Notice that $m_2>m$. It is clear that model I is the limit
of model II when $\deltaw\to 0$ and $m_2\to\infty$ with the product
$m_2\deltaw = \alpha$ finite.

For this model II, the following symmetrization of the problem
explained below will be useful. If an external potential is acting
differently on positive and negative particles, as in our case, in
order to symmetrize the problem for the two types of particles, Cornu
and Jancovici~\cite{CornuJanco} propose to define $ m\left(\r\right)$
and $ V\left( \r\right) $ by the relations
\begin{equation}
\label{m(r)}
m_{s}\left( \r\right) =m\left( \r\right) 
\exp \left[ -2sV\left( \r\right) \right]\,,
\qquad s=\pm 1
\end{equation}
where $V(\r)$ in general is different from the external $V_{\pm}(\r)$
potentials. Knowing that $V_{-}(\r)$ is constant in each region, in
terms of the modified Green functions
\begin{equation}
\label{g(G)}
g_{s_{1}s_{2}}\left( \r_{1},\r_{2}\right) =
e^{-s_{1}V\left( \r_{1}\right)} 
G_{s_{1}s_{2}}\left( \r_{1},\r_{2}\right)
e^{-s_{2}V\left( \r_{2}\right)}
\end{equation}
the system of equations~(\ref{eq:GreenFunctions-G}) can be shown to be
equivalent to
\begin{equation}
\label{eq:delta}
\left[ \left( m(r_{1})\right) ^{2}-\Delta \right] 
g_{\pm \pm }\left( \r_{1},\r_{2}\right) 
= m(r_{1})\delta \left( \r_{1}-\r_{2}\right)
\end{equation}
\begin{equation}
\label{eq:homogenea}
\frac{e^{\pm
i\theta_1}}{m(r_{1})}\left(-\partial_{r_1}\mp\frac{i}{r_1}\partial_{\theta_1}\right)g_{\pm
\pm }\left( \r_{1},\r_{2}\right) = g_{\mp \pm }\left(
\r_{1},\r_{2}\right)
\end{equation}
where $\Delta$ is the Laplacian operator and  
\begin{equation}
m(r)=
\begin{cases}
m_0=(mm_2)^{1/2}
&\text{if $r$ in regions 1 or 3}\,,
\\
m &\text{if $r$ in region 2}\,.
\end{cases}
\end{equation}
and the potential defined in Eq.~(\ref{m(r)}) is given by
\begin{equation}
\exp \left( V(\r)\right)=
\begin{cases}
\left( m_{2}/m\right) ^{1/4}
&\text{if $r$ in regions 1 or 3}\,,
\\
1&\text{if $r$ in region 2}\,.
\end{cases}
\end{equation}
To obtain the Green functions $g_{\pm\pm}$
we solve Eq.~(\ref{eq:delta}) which is a standard inhomogeneous
Helmoltz equation in each region: inner border $R_1<r<R_1+\deltaw$
(region 1), bulk $R_1+\deltaw<r<R_2-\deltaw$ (region 2) and outer border
$R-\deltaw<r<R$ (region3). Outside the annulus $r<R_1$ and $r>R_2$ the
fugacities vanish and the equation satisfied by $G_{\pm \pm}$ is the
Laplace equation.  The Green functions $G_{s_1 s_2}$ are continuous at
each boundary $r=R_1$, $r=R_1+\deltaw$, $r=R_2-\deltaw$ and
$r=R_2$. Remember that here the Green functions $G_{s_1 s_2}$ are
different from the auxiliary functions $g_{s_1 s_2}$. Their
relationship is given by Eq.~(\ref{g(G)}). While the Green functions
$G_{s_1 s_2}$ are continuous, the auxiliary Green functions $g_{s_1
s_2}$ are in general discontinuous at the interfaces.

In the regions of interest, the form of the solution is, in general,
\begin{eqnarray}
\label{eq:form-solut-model-II}
g_{\pm\pm}(\r_1,\r_2)&&= \frac{m(r_1)}{2\pi} \times \\
&&\sum_{l=-\infty}^{+\infty} e^{il(\theta_1-\theta_2)}\left[
I_l(m(r_1)r_<)K_l(m(r_1)r_>)\right.\nonumber\\
&&+\left. B^{\pm}_l(r_2) I_l(m(r_1)r_1) +C^{\pm}_l(r_2) K_l(m(r_1)r_1)
\right]\nonumber
\end{eqnarray}
with $m(r)=m_0$ in the border regions 1 and 3 and $m(r)=m$ in the bulk
region 2. Imposing the continuity of the Green functions $G_{s_1 s_2}$
gives a linear system of equations for the coefficients $B_l^{\pm}$
and $C_l^{\pm}$ which can be conveniently solved with the aid of
matrix algebra software like \textsc{Mathematica}.

\begin{figure}
\includegraphics[width=\GraphicsWidth]{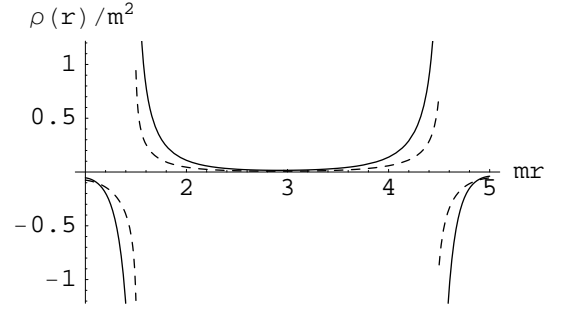}
\caption{
\label{fig:rhototal-step}
The charge density profile for an annulus with inner radius $R_1=1/m$,
outer radius $R_2=5/m$ in model II. The width of the borders are
$\deltaw=0.5 / m$. The fugacity in the border regions is: for the solid
line curve $m_2= 16 m$ corresponding to a value of $m_0=4m$ and for the
dashed line curve $m_2= 4 m$ corresponding to a value of $m_0=2m$. The
cutoff in the sums of Bessel functions in the expression of the
density has been chosen as $N=100$.  }
\end{figure}

The explicit expression of the coefficients obtained from the solution
of the linear systems~\cite{Lina-tesis} are to long to reproduce here,
however we show in Figure~\ref{fig:rhototal-step} a plot of charge
density profile for two values of $m_2$, $m_2=16m$ and $m_2=4m$.

%%%%%%%%%%%%%%%%%%%%%%%%%%%%%%%%%%%%%%%%%%%%%%%%%%%%%%%%%%%%
%
%
%    Figure 8
%    File: rhototal-step.eps
%    Label: fig:rhototal-step
%
%
%
%%%%%%%%%%%%%%%%%%%%%%%%%%%%%%%%%%%%%%%%%%%%%%%%%%%%%%%%%%%%

The discontinuity of the charge density profile near the interfaces
is due to the fact that the density profile of the negative particles
is discontinuous there. This is expected for the negative particles
since the potential $V_{-}(r)$ is discontinuous across the
interfaces. It is clear from our formalism that the density profile
should have the same kind of discontinuity that the Boltzmann factor
of $V_{-}(r)$ since $\rho_{-}(r)=m_{-}(r)\, G_{--}(\r,\r)$ with
$m_{-}(r)=\penalty10000 m \exp(-\beta V_{-}(r))$ and $G_{--}$
continuous at the interfaces. Actually this is a more general result
of statistical mechanics of inhomogeneous fluids at discontinuous
interfaces. In general for a fluid with density $n(x)$ near a planar
interface characterized by an external potential $V_{\text{ext}}(x)$
eventually discontinuous at $x=0$ the $y$-function $\exp(\beta
V_{\text{ext}}(x))\, n(x)$ is continuous~\cite{Henderson}.

There is a higher density of particles (both negative and positive) in
the borders that in the bulk: for higher values of the fugacity $m_2$
in the border (the external potential $V_{-}(\r)$ is more attractive),
the density in the borders is higher. In region 2, the density far
from the interfaces is close to the bulk
value~(\ref{eq:bulk-density}). In the borders, away from the
interfaces the density tries to be close to the new bulk value given
by Eq.~(\ref{eq:bulk-density}) replacing $m$ by $m_0$. If the width
$\deltaw$ of the border region was much larger than $m_0^{-1}$, one
would expect that the density far away from the interfaces converges
to a bulk value according to the value of $m(r)$ in that region.  This
is true for both the negative and positive of particles.  Due to the
natural tendency of the system to be electrically neutral, the
positive particles try to follow the negative ones so the system is
not locally charged. However at the interfaces $r=R_1+\deltaw$ and
$r=R_2-\deltaw$ there remains a non-neutral charge density that can be
seen in Figure~\ref{fig:rhototal-step}. Actually one can see in
Figure~\ref{fig:rhototal-step} a double charged layer. Inside the
border region (1 or 3) there is a negative charge density layer and
outside the border, in the bulk region 2, a positive layer, the same
one that was previously observed with model I in
Figure~\ref{fig:rhoannulus}. These double layers are concentrated near
the interfaces at $r=R_1+\deltaw$ and at $r=R_2-\deltaw$ and have a
thickness of order $m^{-1}$ for the positive layer in region 2 and
$m_0^{-1}$ for the negative layer in regions 1 and 3.

\section{Summary and perspectives}

The present solvable model studied here gave us interesting
information about the behavior of confined Coulomb systems with
attractive boundaries. This system has an induced internal charge on
the boundary which is created by an external potential which is not of
electrical nature. This potential only acts on the negative particles,
while the positive particles are unaffected. This is not the usual
situation that has been studied extensively in the past, where the
system is submitted to electrical forces due to possible external
charges.

First, we found that large systems exhibit the same finite-size
corrections that for systems without attractive boundaries, confirming
again the universal nature of these finite-size corrections. Studying
the disjoining pressure we found that the attractive boundaries have a
stabilizing effect. This was noticed also in our previous
work~\cite{TellezMerchan-jabon}, however the curvature in the present
case is very important. It makes the surface tension to be the
predominant contribution to the disjoining pressure, as opposed to the
slab geometry. Then, we conclude that the curvature has also a
stabilizing effect on the system in comparison to the slab geometry
in which the system can be unstable for low values of the adhesivity.

The study of the density profiles gives information about the
structure of the system. As expected, there are some adsorbed charges
on the boundary and these are screened by a positive layer of charge
inside the system. We were able to check explicitly an
electro-neutrality sum rule and a few relations that the adsorbed
charge in the boundary satisfy.

It would be interesting to know what features of the present model are
universal and which are not. A step toward answering this question can
be obtained by studying another solvable model of Coulomb system, the
one-component plasma. A preliminary study of this system was done in
Ref.~\cite{Lina-tesis} and this will be the subject of a future paper.

\begin{acknowledgments}
The authors would like to thank the following agencies for their
financial support: ECOS-Nord (France), COLCIENCIAS-ICFES-ICETEX
(Colombia), Banco de la Rep\'ublica (Colombia) and Fondo de
Investigaciones de la Facultad de Ciencias de la Universidad de los
Andes (Colombia).
\end{acknowledgments}

%%%%%%%%%%%%%%%%%%%%%%%%%%%%%%%%%%%%%%%%%%%%%%%%%%%%%%%%%%%%%%%

\end{document}